\documentclass[aps,pre,10pt,twocolumn,amsfonts,amssymb,amsmath,superscriptaddress]{revtex4-1}
\usepackage{graphicx}
\usepackage[colorlinks=true,linkcolor=blue,citecolor=blue]{hyperref}	
\usepackage{dcolumn}
\usepackage{bm}
\usepackage{subfigure}

\usepackage{wasysym}

\newcommand{\Tc}{\ensuremath{T_\mathrm{c}}}   
\newcommand{\pc}{\ensuremath{p_\mathrm{c}}}   
\newcommand{\pd}{\ensuremath{p_\mathrm{d}}}   
\newcommand{\rc}{{\ensuremath{r_\mathrm{c}}}} 

\newcommand{\red}[0]{\textcolor{red}{}}

\begin{document}



\title{Robustness of functional networks at criticality against structural defects}



\author{Abdorreza Goodarzinick}
\affiliation{Department of Physics, Institute for Advanced Studies in Basic Sciences (IASBS), Zanjan 45137-66731, Iran}

\author{Mohammad D. Niry}
\email{m.d.niry@iasbs.ac.ir}
\affiliation{Department of Physics, Institute for Advanced Studies in Basic Sciences (IASBS), Zanjan 45137-66731, Iran}

\author{Alireza Valizadeh}
\affiliation{Department of Physics, Institute for Advanced Studies in Basic Sciences (IASBS), Zanjan 45137-66731, Iran}
\affiliation{School of Cognitive Sciences, Institute for Research in Fundamental Sciences (IPM), P.O. Box 1954851167, Tehran, Iran}

\author{Matja{\v z} Perc}
\affiliation{Faculty of Natural Sciences and Mathematics, University of Maribor, Koro{\v s}ka cesta 160, SI-2000 Maribor, Slovenia}
\affiliation{CAMTP -- Center for Applied Mathematics and Theoretical Physics, University of Maribor, Mladinska 3, SI-2000 Maribor, Slovenia}
\affiliation{School of Electronic and Information Engineering, Beihang University, Beijing 100191, P.R. China}


\date{\today}

\begin{abstract}
The robustness of dynamical properties of neuronal networks against structural damages is a central problem in computational and experimental neuroscience. Research has shown that the cortical network of a healthy brain works near a critical state, and moreover, that functional neuronal networks often have scale-free and small-world properties. In this work, we study how the robustness of simple functional networks at criticality is affected by structural defects. In particular, we consider a 2D Ising model at the critical temperature and investigate how its functional network changes with the increasing degree of structural defects. We show that the scale-free and small-world properties of the functional network at criticality are robust against large degrees of structural lesions while the system remains below the percolation limit. Although the Ising model is only a conceptual description of a two-state neuron, our research reveals fundamental robustness properties of functional networks derived from classical statistical mechanics models.
\end{abstract}


\maketitle


\section{Introduction}
The brain as a complex system, consists of tens of billions of highly nonlinear components and exhibits collective dynamics which in many aspects resemble some of the well-known phenomena in statistical physics. Many researchers believe that the emergence of functions, such as consciousness, learning and memory in face of the real world requires a critical dynamics in the brain \cite{Chialvo2010,schuster2014criticality}. In self-organized critical systems, the perturbations can die out in very short distances or propagate through the network as avalanches with no characteristic time and length scale, indicating the system is in critical state \cite{Bak1987}. Unlike the conventional critical states in equilibrium statistical mechanics which occur with fine-tuned parameters, these critical states are attractors reached by starting far from equilibrium; they are insensitive to parameters and many dynamical systems with extended spatial degrees of freedom evolve into such critical states \cite{Bak1987,moosavi2017refractory}. Over recent years, there has been growing evidence supporting the hypothesis that the cortex acts as a system near it's critical point \cite{Chialvo2010,hesse2014self,shew2015adaptation}. Amplitude of neural oscillations' fluctuations show long-range correlations and power-law scaling behavior \cite{Linkenkaer-hansen2001}. Considering the small scale, the successive activity of ensembles of neurons known as neuronal avalanches obey power-law distributions in both size and duration \cite{Beggs2003}. In the large scale, the brain's functional network, which its edges represent correlation between voxels in fMRI image correspondingly obtained from human brain, are known to be scale-free and small-world networks, along with power-law decay of correlations over the distance \red{\cite{Eguiluz2005,buzsaki2014log,haimovici2013,expert2011,fraiman2012}}. The functional advantages of being in the critical state may further support this hypothesis \cite{Shew2013,massobrio2015criticality}.

Experience shows that the brain is robust against many kinds of damage. Several degenerative neurological disorders such as Alzheimer's, Parkinson's, and Huntington's, show no functional symptoms and remain indistinguishable by simple clinical diagnosis for many years \cite{sperling2014evolution,pievani2014brain}. The general view is that this resistance is due to the resilience of complex networks (structural neural network) to destruction. But the question is that: Is the critical dynamics of the brain contributing to this robustness? According to the authors knowledge two simple models have been studied whose functional networks are similar to the functional network of the brain \cite{Fraiman2009, Zarepour2015}. In the current paper, we select the Ising model and by adding defects, investigate the effect of lessions on the derived functional network.
\red{Presence of defects can produce structural heterogeneity in the network which can change the standard single critical-point in the case of regular systems, to a broadened range of parameters with power-law scaling and the so-called Griffiths phase which has been seen in hierarchical modular networks in synthetic and biological networks \cite{griffiths1969,munoz2010, moretti2013}. Although a widened range of parameters within which the system can retain criticality may have functional advantages, it is not seen in our model since we have focused on the dynamics of the spanning cluster by discarding the isolated regions when extracting simulation data.}


\section{\label{sec:meth} Methods}
Ising model \cite{Ising1925} is widely used as a prototypical model for studying the cortical phenomena \cite{roudi2009ising,kadirvelu2017inferring,marinazzo2014information}. We have used a 2D Ising model for a configuration of the spins $\left\{ \sigma_{i} \right\}$ described by the Hamiltonian
\begin{eqnarray}\label{eq.1}
H(\left\{ \sigma_{i}\right\}) = -J\sum_{[i,j]}^{N} \sigma_{i} \sigma_{j},
\end{eqnarray}
where the summation is taken over all the nearest neighbors and $J$ is the interaction strength. In the rest of paper, for simplicity, the model is considered in a reduced unit system where the energy and the temperature are measured in the units of $J$ and $J/k_\mathrm{B}$, respectively, in which $k_\mathrm{B}$ represents the Boltzmann constant.

We implemented the Metropolis Monte Carlo algorithm to study the evolution of the spins \cite{Metropolis1953}. At each Monte Carlo step all the $N=L^2$ spins are given a chance to flip. For each simulation result, the expected value of statistical quantities are calculated over $10^3$ Monte Carlo steps by using $4$ to $10$ distinct realizations to obtain the desired accuracy. At the critical point, scaling behavior is observed in multiple thermodynamic quantities. Although the critical temperature itself, depends on the structure and details of interaction, scaling behavior is not affected by these and only depends on some fundamental geometrical properties such as the system's dimension \cite{Kardar2007}.

Functional networks are defined based on correlations between the activity of the nodes (i.e. either a ``voxel" in case of the brain \cite{Eguiluz2005}, a lattice site in the Ising model \cite{Fraiman2009}, or in the Abelian sandpile model \cite{Zarepour2015}). Here, we have used the Pearson's correlation coefficient defined as:
\begin{eqnarray}
\label{eq.2}
r(i,j)=\frac{\langle x_{i}(t)x_{j}(t) \rangle - \langle x_{i}(t) \rangle \langle x_{j}(t) \rangle}{\sigma[x_i(t)] \sigma[x_j(t)]},
\end{eqnarray}
where $\langle\cdots\rangle$ represents average taken over the length of the time series and $\sigma [\cdots]$ is the standard deviation of the activity at the corresponding site. Two nodes $i$ and $j$ are functionally connected if the correlation $r(i,j)$ is equal or larger than a predetermined constant value $\rc$. In the current paper the results are presented for $\rc=0.4$, while other values (\rc=0.3, 0.5) were also tested and the results were qualitatively the same \red{(see the discussion in the last paragraph of Section \ref{sec:res})}.

\begin{figure}
\subfigure[]{\includegraphics[height=0.22\columnwidth,angle=90]{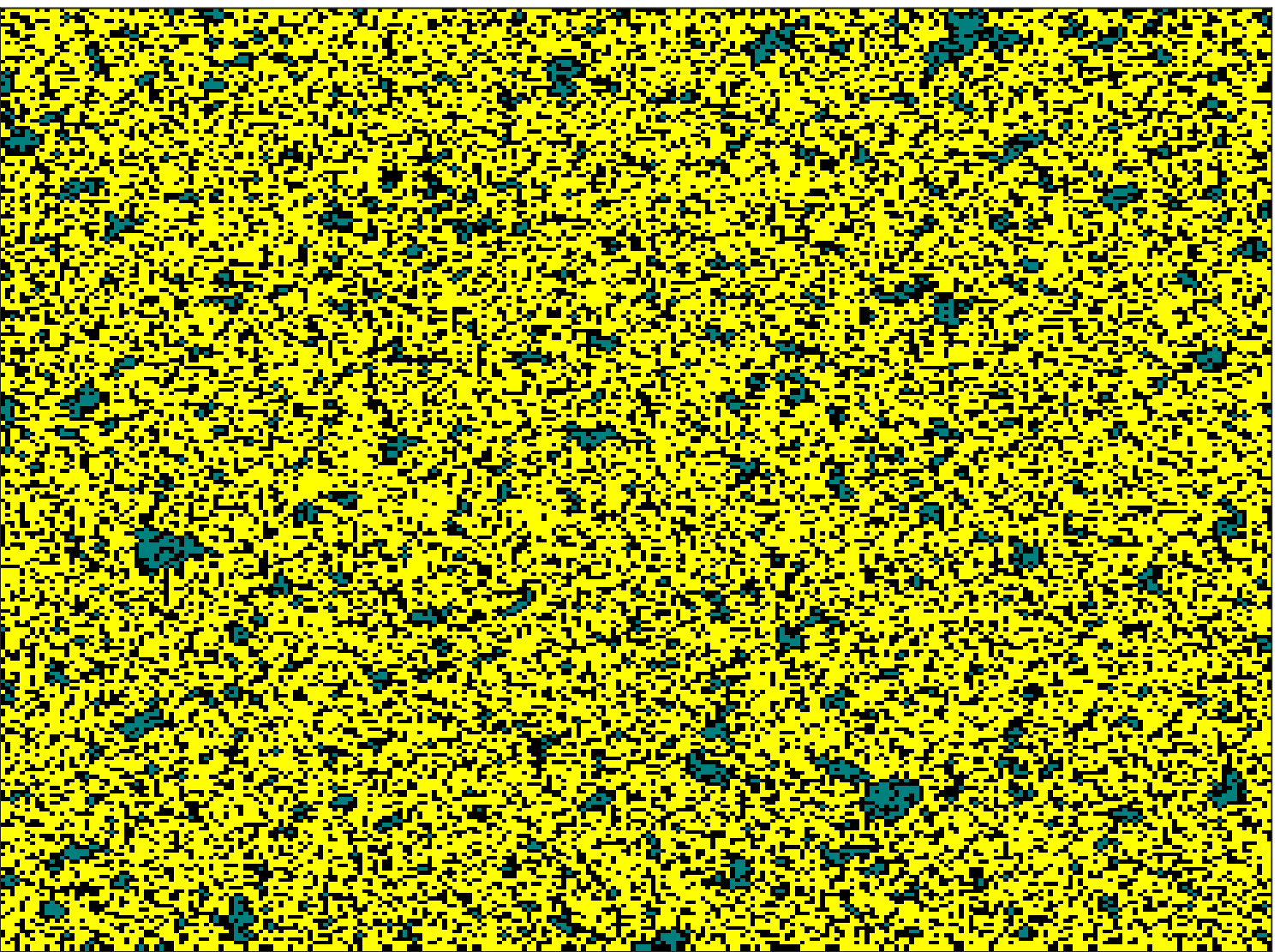}}
\hfil
\subfigure[]{\includegraphics[height=0.22\columnwidth,angle=90]{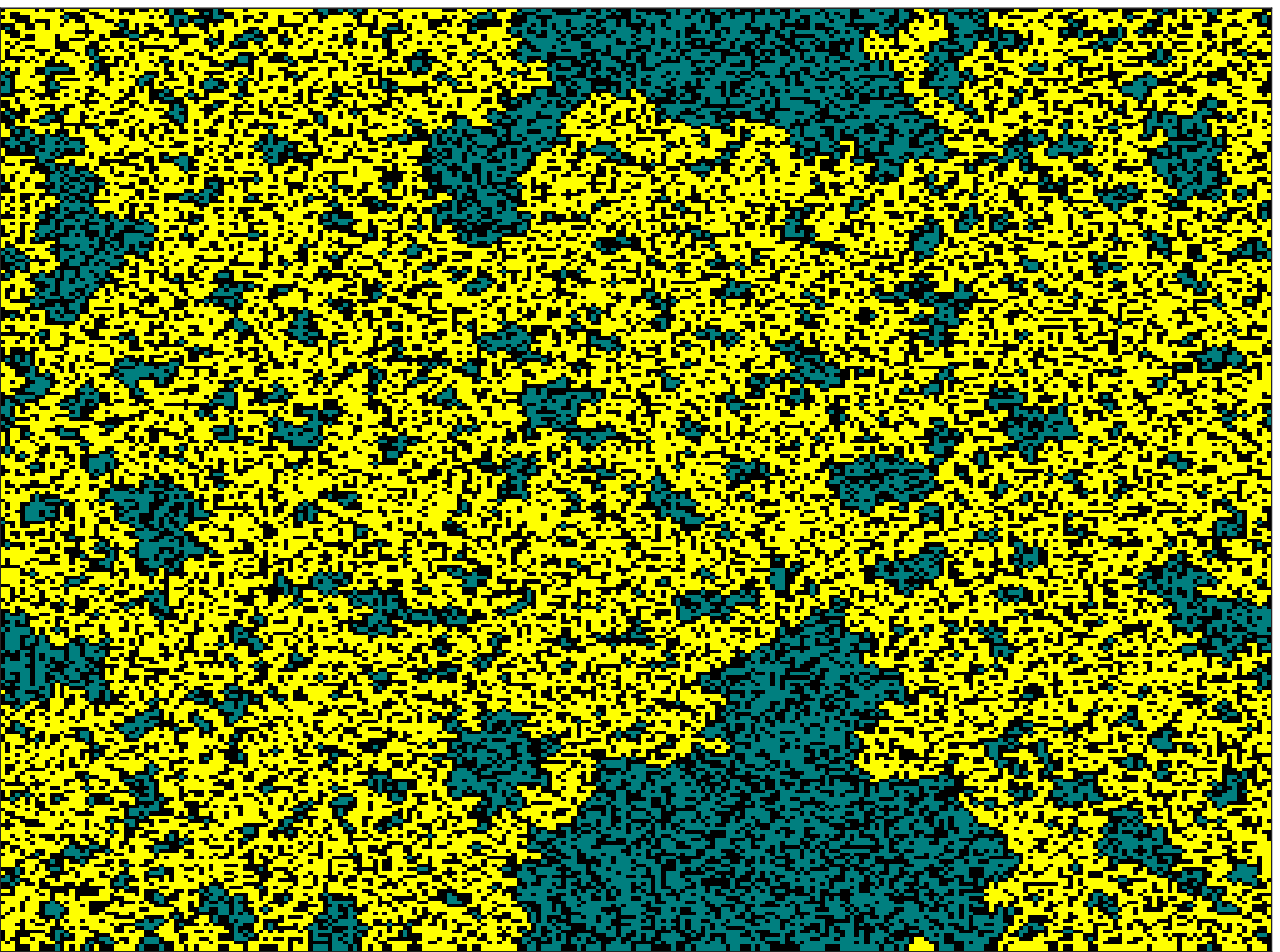}}
\hfil
\subfigure[]{\includegraphics[height=0.22\columnwidth,angle=90]{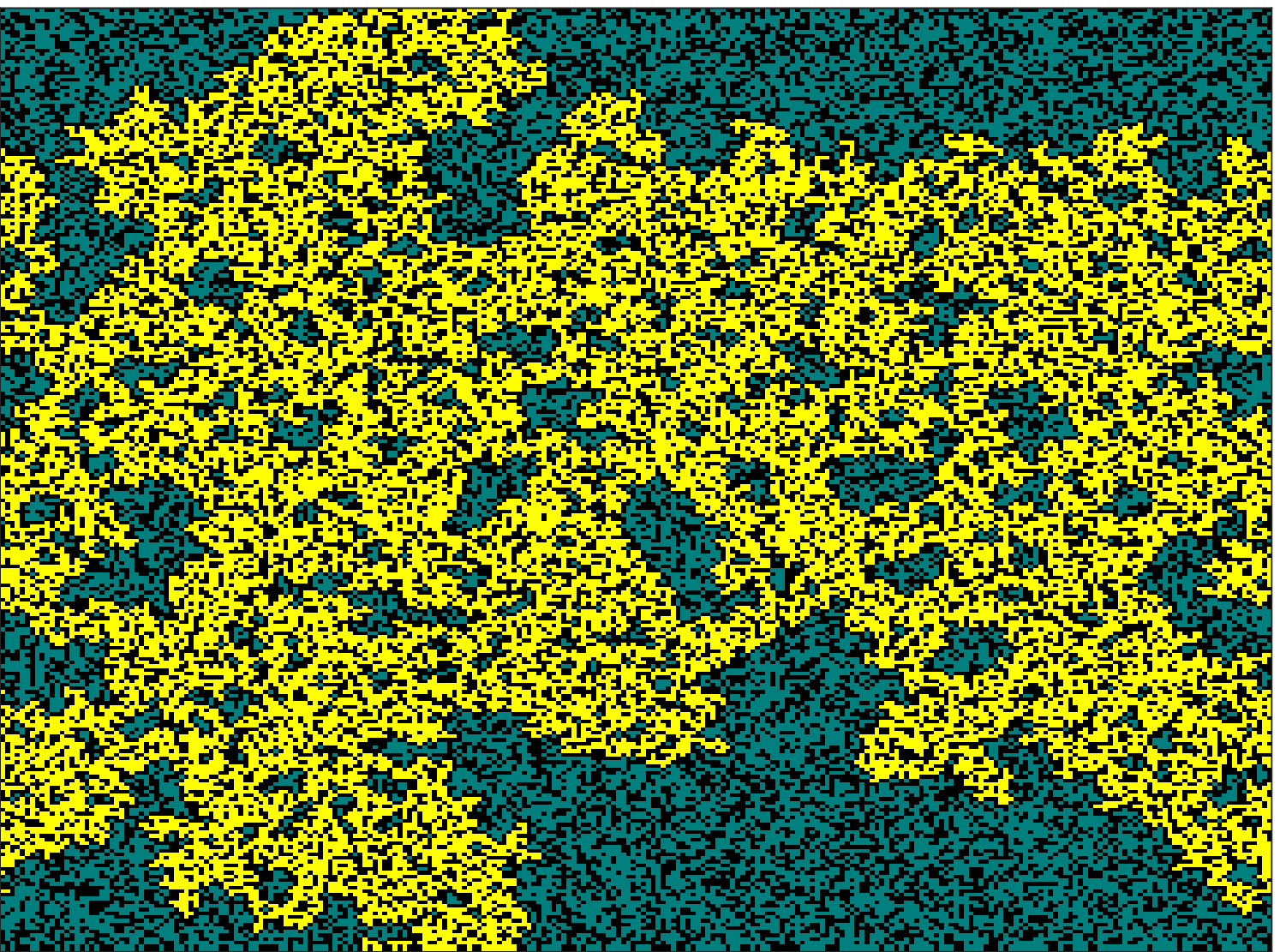}}
\hfil
\subfigure[]{\includegraphics[height=0.22\columnwidth,angle=90]{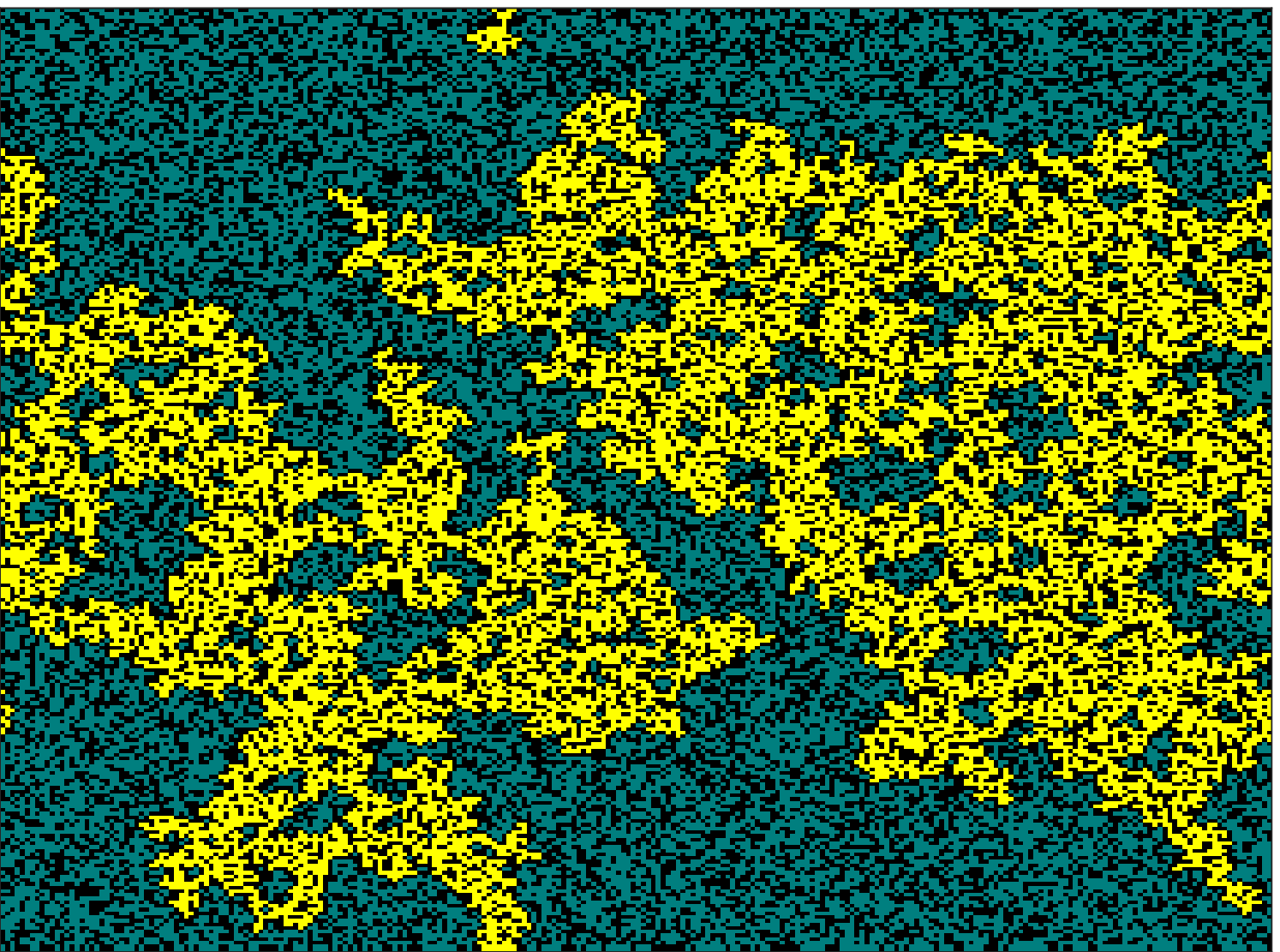}}
\caption{\label{fig:HoshenKopelman} (color online) Defect ratio $\pd$ in the lattice is set $0.35$, $0.40$, $0.41$, and $0.42$ from (a) to (d). Light color shows intact cells, while main defected cells are indicated by black. Also, isolated regions which were identified and removed by the Hoshen-Kopelman algorithm are dark shaded.}
\end{figure}

We have introduced defects by choosing a $\pd$ fraction of sites randomly, and setting their spin to zero, thus deactivating them and removing their interactions with up to four nearest neighbors. Moreover, increasing the fraction of defects, some areas appear in the lattice which are completely surrounded by deactivated sites, so they can not interact with the spanning cluster of active spins. \red{These isolated clusters act like decoupled Ising ``sub-systems" having super-critical dynamics at the critical temperature of the main network (i.e., the spanning cluster) \cite{griffiths1969}, and their functional network will be independent from that of the giant cluster and shows Poisson statistics \cite{Fraiman2009}.} This issue inflicts errors in computing correlations and the derivation of the functional network. To deal with this problem and to improve computations, we used the algorithm proposed by Hoshen and Kopelman to identify the spanning cluster \cite{HoshenR.Kopelman1976}, only considering the giant cluster to extract data, excluding isolated clusters (e.x., Fig.~\ref{fig:HoshenKopelman}). The procedure of the deactivation of the sites can be continued while the set of active spins (a fraction of $1-\pd$ of lattice sites) percolates and there exists a giant cluster of active spins with the diameter of the order of system size $L$ which touches boundaries of the lattice \cite{Newman2000}.

\begin{figure}
	\includegraphics[width=0.9\columnwidth]{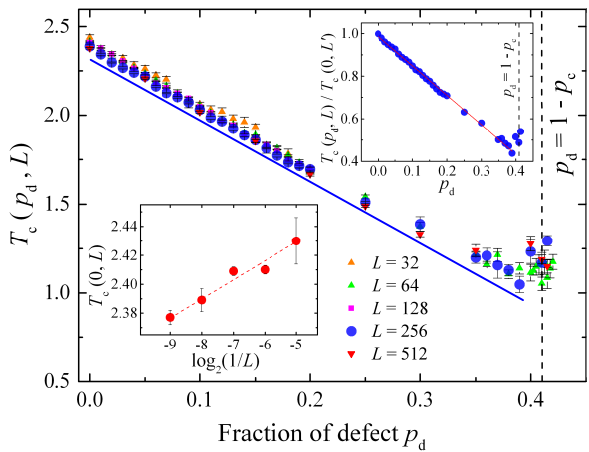}
	\caption{\label{fig:Tc} (color online) Average calculated critical temperature at every stage of elimination of structural sites. The error bars show the standard error over the ensemble average. Vertical dashed lines show the percolation limit $\pd = 1-\pc \simeq 0.41$. Thick straight line is linear fit, where its the intercept relates to $L$ as was shown in the lower inset, so $\Tc(\pd, L) \cong \Tc(0, L) -3.45 \pd$. The upper inset shows the normalized critical temperature where number of intact spins $L'^2 \approx (1-\pd)L^2$ and the straight line is linear fit to the data points as $\Tc(\pd,L) / \Tc(0,L^\prime) = 1.0 - 1.39 \pd$. Regression coefficients for all the linear fits are above $0.98$.}
\end{figure}

Beginning with a perfect two dimensional lattice, we first found the critical temperature at which the constant-volume specific heat is maximized. When defects are put into the system, the size and the topology of the system will not be the same as the intact system and both may contribute in changing the critical temperature. At each level of the lesion characterized by the fraction of defected sites $p_d$, we re-measured the critical temperature $\Tc(\pd, L)$ and adjusted the temperature to retain the system's criticality. The dependence of the critical temperature on the lesion degree shows a quasi-linear trend almost up to the percolation limit $p_d \simeq 0.41$ (Fig.~\ref{fig:Tc}). \red{It is worth mentioning that the decomposition of the system into smaller clusters leads to the well-known Griffiths phase and stretchig of the scaling region \cite{griffiths1969,vojta2006rare}. So, when the isolated clusters are removed, the only surviving phase at $\Tc(\pd, L)\le\Tc(0, L)$ will be the conventional ferromagnetic phase.} The reliability of the results is tested by changing the lattice size from $L = 32$ to $512$ and observing a linear relation between critical temperature and the lattice diameter in a semi-log scale (Fig.~\ref{fig:Tc} lower inset). The normalized critical temperature is plotted vs $\pd$ in the upper inset of Fig.~\ref{fig:Tc}. It shows the slope of quasi-linear trend of the critical temperature slightly depends on the effective size of the lattice, $L'$, which is obtained from the square root of the number of intact spins. For clarity, this inset is only plotted for the specific lattice size $L = 2^8$ which is also used to obtain most of our results explained in the next section.\\

\begin{figure}
	\definecolor{darkorange}{RGB}{235, 23, 23}
	\definecolor{darkgreen}{RGB}{53, 74, 87}
	\definecolor{orange}{RGB}{255, 150, 0}
	\subfigure[]{\label{fig:N(a)}\includegraphics[width=0.48\columnwidth]{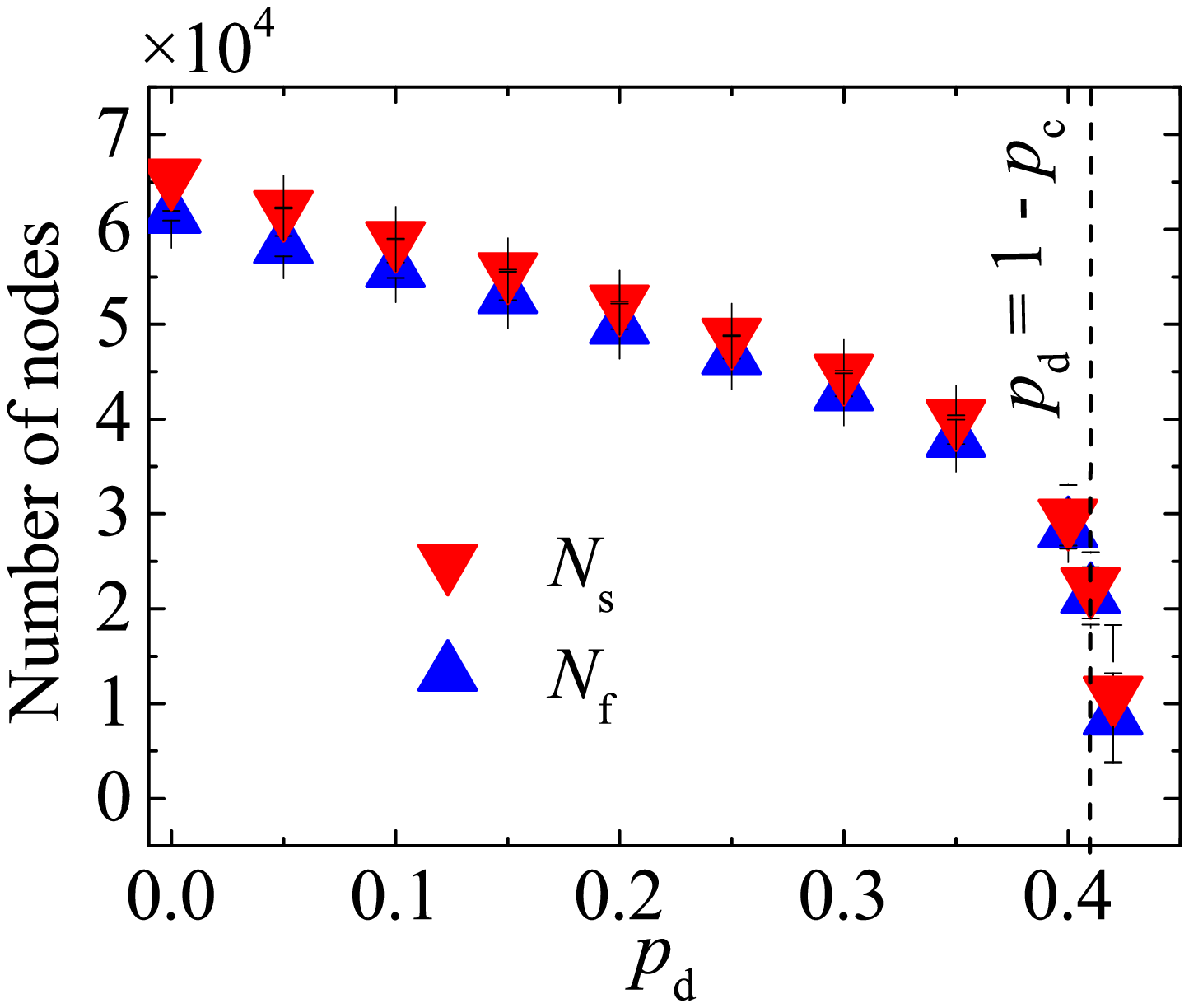}}
	\hfill
	\subfigure[]{\label{fig:N(b)}\includegraphics[width=0.48\columnwidth]{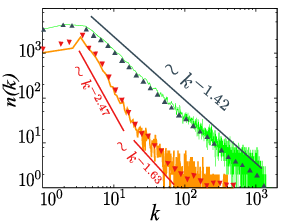}}
	\caption{\label{fig:N} (color online) (a) depicts the number of nodes in the network vs the amount of defects $\pd$, up to the percolation limit. (\textcolor{red}{$\blacktriangledown$}) symbols show the number of nodes in the structural network, $N_\mathrm{s}$, and (\textcolor{blue}{$\blacktriangle$}) symbols show the number of nodes in the functional networks, $N_\mathrm{f}$. (b) is an example of degree-distributions of scale-free functional networks obtained before (after) removal of isolated clusters at $\pd=0.35$ shown by \textcolor{orange}{thick} (\textcolor{green}{thin}) curves. To improve clarity of the graph and decrease the error caused by fluctuations, linear regressions are carried out on logarithmic averages shown by \textcolor{darkorange}{$\blacktriangledown$} (\textcolor{darkgreen}{$\blacktriangle$}) symbols. Straight thick lines represent the power-law fit over the average results.}
\end{figure}

As shown in Fig.~\ref{fig:N(a)}, the number of nodes in the structural networks is always higher than number of nodes in the functional networks and both of them decrease with increasing the amount of defects. At the critical temperature, the functional network derived from the 2D Ising model becomes scale-free and small-world, while away from the critical point, the network is a locally-connected random one \cite{Fraiman2009}. So to characterize the functional network, we have measured the most relevant network characteristic properties, the scaling exponent of degree-distribution $\gamma$, and small-worldness measure $S$. The scaling exponent was calculated using linear fit to the log-log data of log-binned degree distribution function of the functional network with exclusion of lower cut-off region [Triangles in Fig.~\ref{fig:N(b)}]. All parameters required to compute $S$ were obtained by using SNAP package Ver.~$3.2$ \cite{leskovec2016snap}. Also Fig.~\ref{fig:N(b)} shows that by removing the isolated spin clusters, as explained in Fig.~\ref{fig:HoshenKopelman}, the crossover in the power-law behavior of degree-distribution function is disappeared and the region of scaling expands.

\section{\label{sec:res} Results and Discussion}
We first analyzed degree distribution of the nodes in the functional network of the Ising model for different degrees of lesion. In \red{Fig.~\ref{fig:N(b)}} an exemplar result is shown for the two cases: when all the active sites are included in the statistics and when only the giant cluster is considered. For a large fraction of defect size near the percolation limit, power-law distribution can be seen only if the isolated clusters are excluded. So hereafter all the results are shown for the spanning cluster. As depicted in \red{Fig.~\ref{fig:N(a)}}, the number of nodes in the giant cluster of the structural network $N_\mathrm{s}$ shows the familiar percolation behavior with a sudden drop near the percolation limit.

\begin{figure}
	\centering
	\subfigure[]{\label{fig:N,E(a)}\includegraphics[width=0.9\columnwidth]{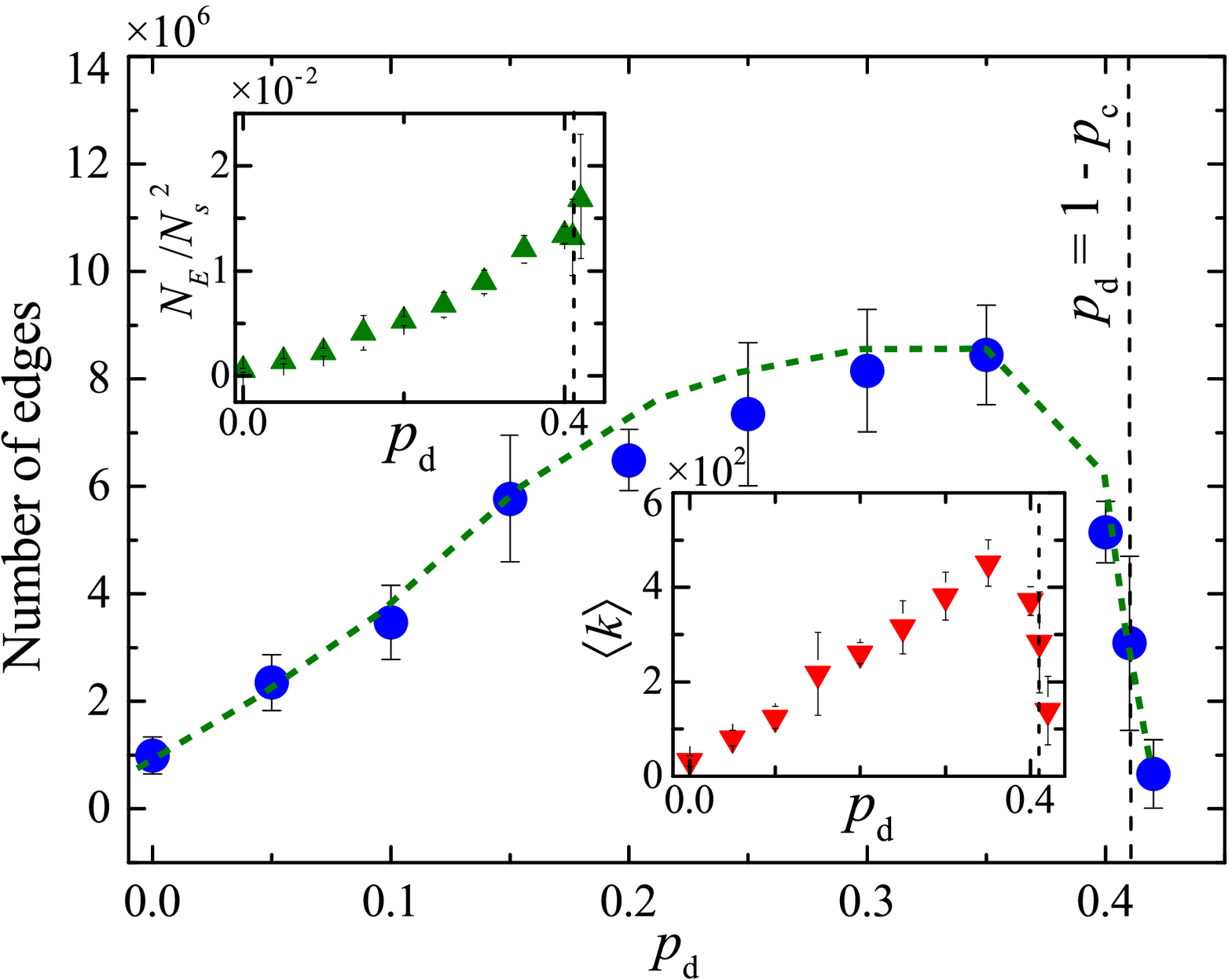}}
	\subfigure[]{\label{fig:N,E(b)}\includegraphics[width=0.9\columnwidth]{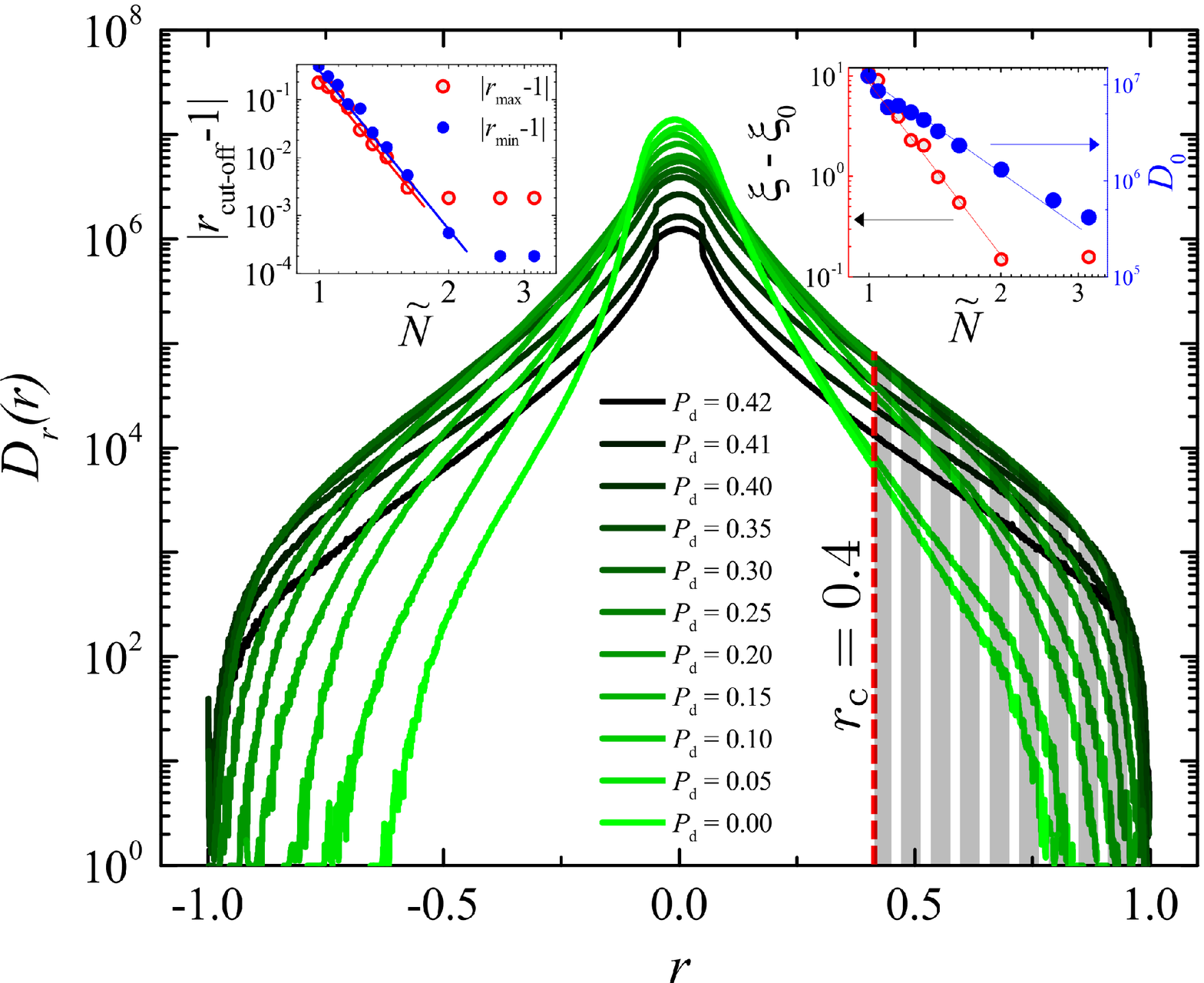}}
	\caption{\label{fig:N,E} (color online) (a) Total number of edges in the functional networks in term of defect ratio up to the percolation limit. Upper and lower insets show the normalized number of edges and average degree, respectively. The standard error over multiple realizations is shown by error bars. The dashed curve shows the estimation based on our model. (b) Distribution of all two-point correlations in the Ising model for different defect degrees. Insets show power-law fits on the parameters in Eq.~(\ref{eq:Nr}). The area of shaded region represents $N_\mathrm{E}$.}
\end{figure}

We also checked the total number of edges in the functional network, i.e., the number of links across which the correlation is above $\rc$. Interestingly, number of edges $N_\mathrm{E}$ and the average node degree $2N_\mathrm{E}/N_\mathrm{f}$ in the functional network increase with lesion size [Fig.~\ref{fig:N,E(a)}] despite to the decrease in the size of structural network [\red{Fig.~\ref{fig:N(a)}}]. Near the percolation limit (at $\pd \simeq 0.35$), both the measures maximize and decrease with further increasing the lesion size. We then evaluated the normalized number of edges in the functional network defined as the number of edges $N_\mathrm{E}$ divided by the total possible number of links for each value of the lesion size. This parameter shows a monotonic increase in the full range of defect size before percolation limit [see the upper inset in Fig.~\ref{fig:N,E(a)}]. These results suggest that the initial increase of the number of edges in the functional network is related to the average increase in the correlation between the nodes, and the subsequent decrease is due to the drop in the size of structural network near the percolation limit. To see how the average correlation between the nodes increases with lesion size, we evaluated the distribution of the correlations between every two nodes in the network for different sizes of the lesion [Fig.~\ref{fig:N,E(b)}]. The results show that with increasing the fraction of defects, the distribution of the correlations widens and maximum correlation increases and saturates to one for $\pd \simeq 0.35$ [see the left inset in Fig.~\ref{fig:N,E(b)} and knowing that $\tilde{N} \approx 2$ for $\pd = 0.35$] which results in the increase in the area under the distribution function curve over which the correlation is larger than the predefined threshold $\rc$. Increasing $\pd$ beyond this range decreases the number of edges with correlation $\rc<r<r_\mathrm{max}$, leading to decrease of the area under the curve in this range, and hence decrease in the number of links in functional network. Quantitatively, we observed an exponential decrease in the non-normalized distribution function of positive and negative correlations with upper and lower cut-offs $r_\mathrm{max}$ and $r_\mathrm{min}$ where decay in the positive region obeys
\begin{eqnarray}\label{eq:Nr}
	D(r) = D_0\exp(-\xi r),
\end{eqnarray}
while coefficients $D_0$, $\xi$, and the cut-offs showed power-law behavior versus the inversed reduced number of nodes $\tilde{N} = L^2/N_\mathrm{s}$ which is shown in the insets of Fig.~\ref{fig:N,E(b)}. By determining $D_0$, as $D(r)dr$ represents the total number of two-point correlations between spin pairs in the range of $r$ and $r+dr$ for specific lattice size $L$, we could calculate the area under the distribution function curves from $\rc$ to $r_\mathrm{max}$, i.e. $\int_\rc^{r_\mathrm{max}}D(r)dr$ which is equal to $N_\mathrm{E}$ and is illustrated by the shaded area in Fig.~\ref{fig:N,E(b)}. This way, we could predict the behavior of $N_\mathrm{E}$ for different values of $\rc$ and $\pd$. The dashed curve in Fig.~\ref{fig:N,E(a)} is the result of this estimation for $\rc = 0.4$.

\begin{figure}
	\includegraphics[width=0.9\columnwidth]{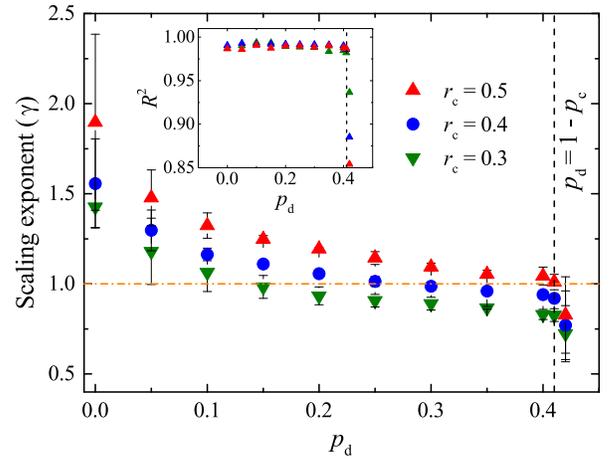}
	\caption{\label{fig:gamma}(color online) Calculated scaling exponent of degree-distributions of the functional network of the Ising model vs. the fraction of defects to the structure up to the percolation limit for three $\rc$ values $0.3$, $0.4$ and $0.5$. The error bars show standard error over multiple realizations. The horizontal dash-dotted line shows $\gamma=1$. Inset shows corresponding R-squared goodness of power-law fits over network data.}
\end{figure}

The degree-distribution of the functional network for giant cluster is checked to obey a power-law distribution, and it is found that up to the percolation point ($\pd\sim0.41$), the best fit for the distributions is a power-law with the widest range of validity among possible regressions. As is shown in inset of \red{Fig.~\ref{fig:gamma}}, R-squared goodness of power-law fit do not show any significant change over this range and the functional network at the critical temperature remains scale-free despite to the lesions. However, the value of the exponent decreases with increasing fraction of defected sites until it saturates around $\pd\approx0.35$. The exponent also depends on the threshold value $\rc$ and increasing $\rc$ raises the $\gamma$-exponent as is shown in Fig.~\ref{fig:gamma}.

Although the saturation at the tail of graph has happened when $\gamma<1$, it can be shown that in the thermodynamics limit the exponent can not be smaller than one. Calculating the normalization constant $Z^{-1}$ for the arbitrary power-law distribution function $p(k) \propto k^{-\gamma}$ as
\begin{eqnarray}\label{eq:Z}
Z &=& \int_1^{k_\mathrm{max}} p(k) dk\nonumber\\[2mm]
&=& (1-\gamma)^{-1} (k_\mathrm{max}^{1-\gamma} - 1),
\end{eqnarray}
yields that $Z$ remains finite when $k_\mathrm{max}\rightarrow\infty$ in the thermodynamics limit, only if $\gamma>1$. Therefore, it can be inferred that for an infinite network, saturation occurs at $\gamma=1$.

\begin{figure*}
	\definecolor{darkgreen}{RGB}{62, 107, 62}
	\begin{minipage}{\textwidth}
			 \subfigure[]{\label{fig:NetParams(a)}\includegraphics[width=0.32\columnwidth]{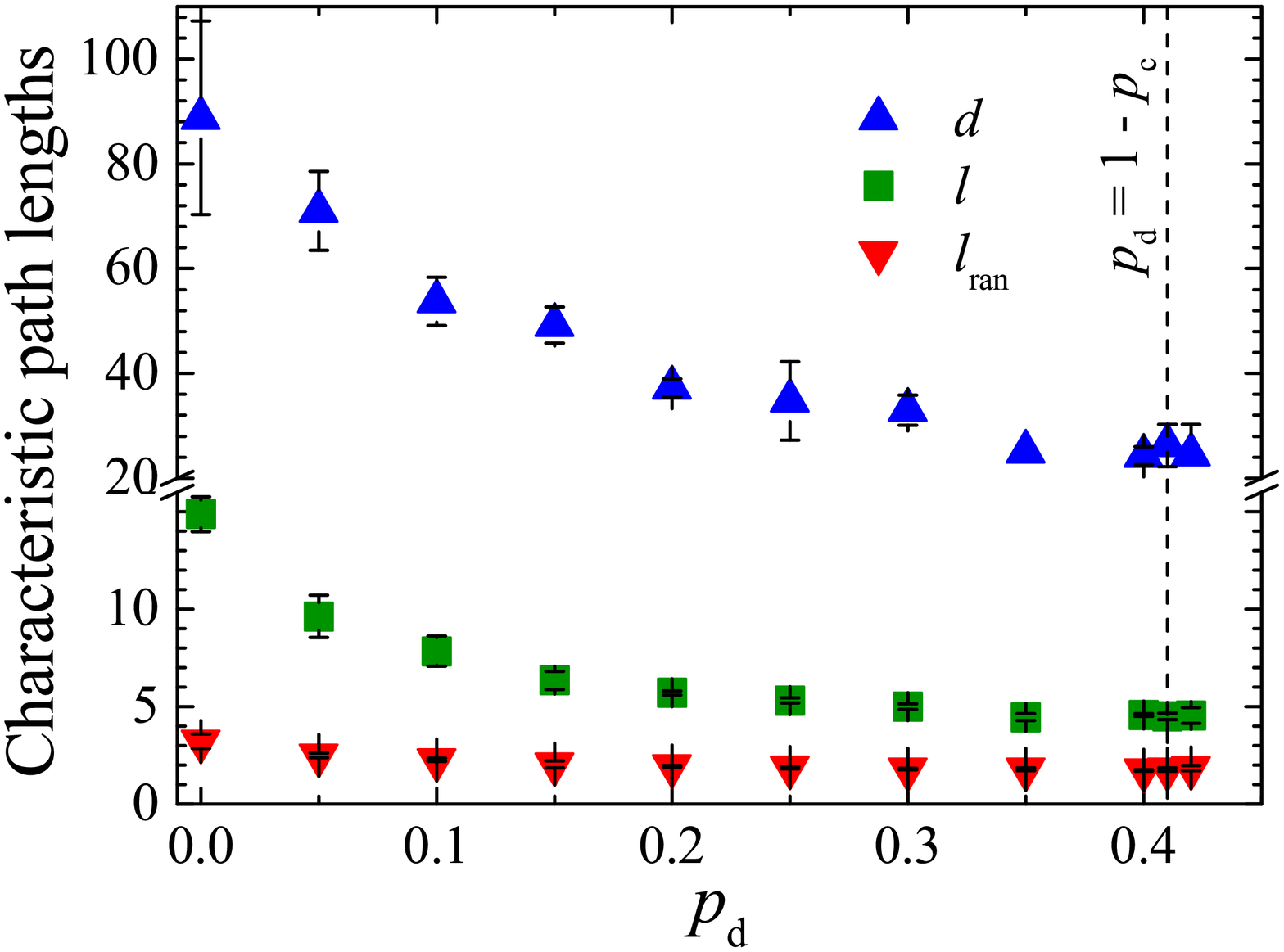}}
			\hfill
			 \subfigure[]{\label{fig:NetParams(b)}\includegraphics[width=0.32\columnwidth]{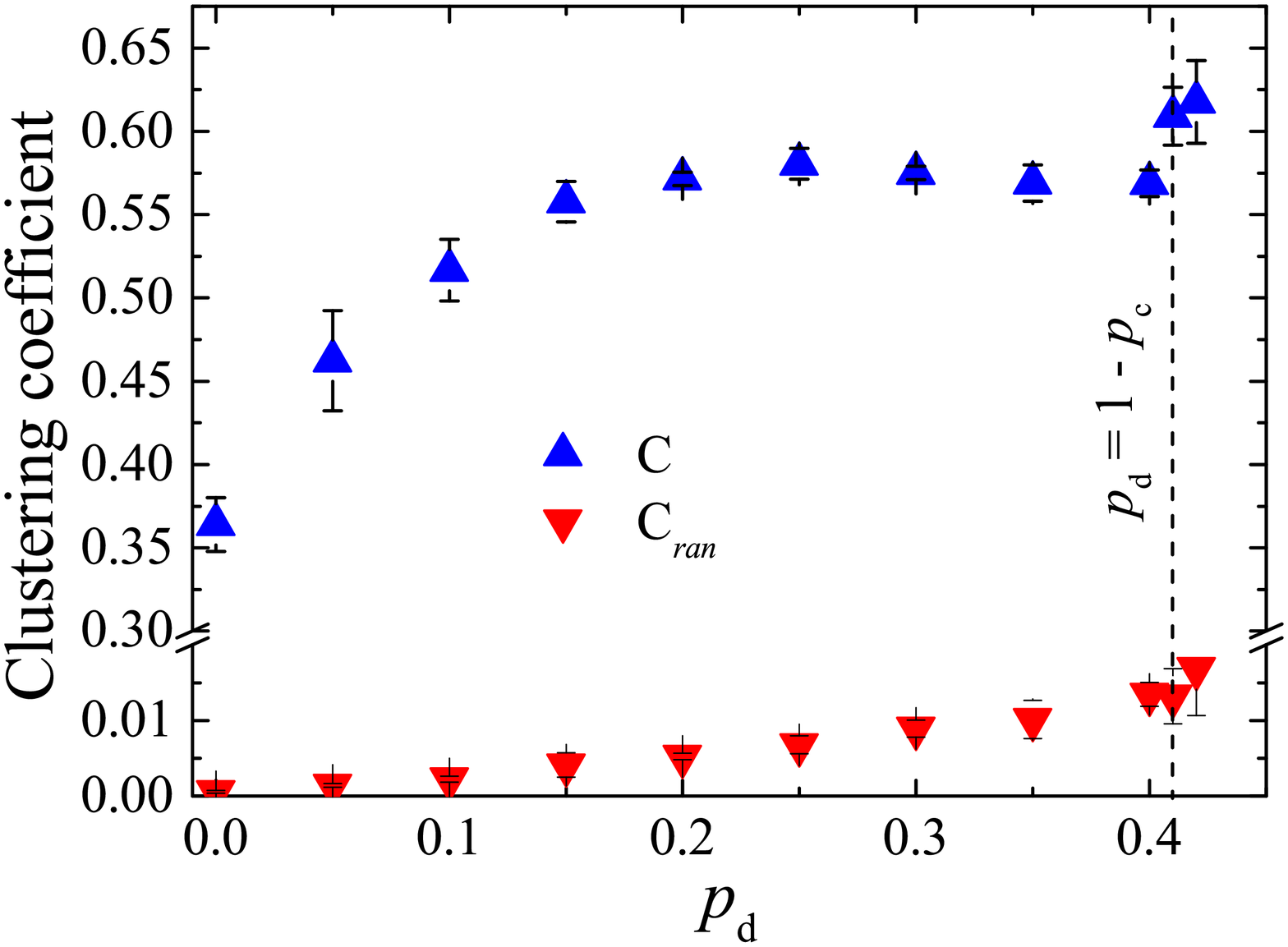}}
			\hfill
			 \subfigure[]{\label{fig:NetParams(c)}\includegraphics[width=0.32\columnwidth]{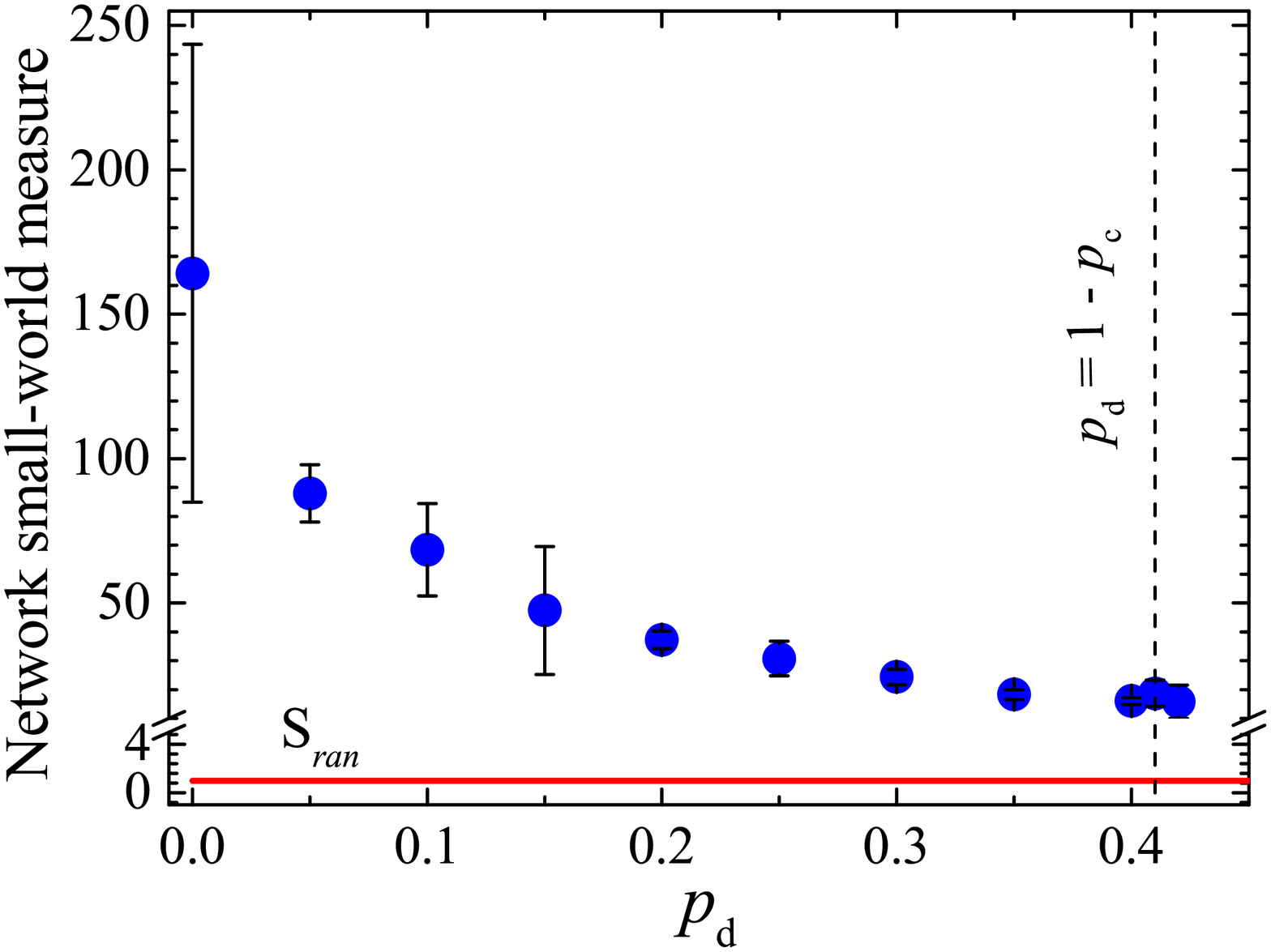}}
	\end{minipage}
	\caption{\label{fig:NetParams} (color online) (a) \textcolor{blue}{$\blacktriangle$} symbols and \textcolor{darkgreen}{$\blacksquare$} symbols show the network diameter $d$ and the average shortest path length $l$ of the functional network for every fraction of defect, respectively. Moreover, the average path length for a corresponding random network $l_\mathrm{ran}$ is represented by \textcolor{red}{$\blacktriangledown$} symbols. (b) Average clustering coefficient of the functional network $C$, and that of a corresponding random network $C_\mathrm{ran}$. (c) Small-worldness measure is calculated according to Eq.~(\ref{eq.3}) vs. $\pd$. Error bars show the standard error over multiple realizations.}
\end{figure*}

As was normalization constant calculated the average degree of vertices can be estimated as
\begin{eqnarray}\label{eq:av_k}
\langle k \rangle &=& \dfrac{\int_1^{k_\mathrm{max}} p(k) kdk}{\int_1^{k_\mathrm{max}} p(k) dk}\nonumber\\[2mm]
	&=& \left(\dfrac{1-\gamma}{2-\gamma}\right) \dfrac{k_\mathrm{max}^{2-\gamma} - 1}{k_\mathrm{max}^{1-\gamma} - 1},
\end{eqnarray}
for $1<\gamma<2$. Also, $\lim_{k_\mathrm{max}\rightarrow\infty} k_\mathrm{max}^{1-\gamma} = 0$ in the same condition. So, the average degree of vertices yields:
\begin{equation}
\langle k \rangle \propto k_\mathrm{max}^{2-\gamma}.
\end{equation}
Therefore, in agreement with results of Fig.~\ref{fig:N,E} and Fig.~\ref{fig:gamma} while $\pd\lesssim0.3$, decrease in the scaling exponent $\gamma$ leads to increase in $\langle k \rangle$. The sudden decrease in $\langle k \rangle$ after $\pd\sim0.35$ is due to the fast decrease in the number of nodes in the spanning cluster of the structural network close to the percolation limit [Fig.~\ref{fig:N,E(a)}].

\begin{figure}[b]
	\includegraphics[width=0.85\columnwidth]{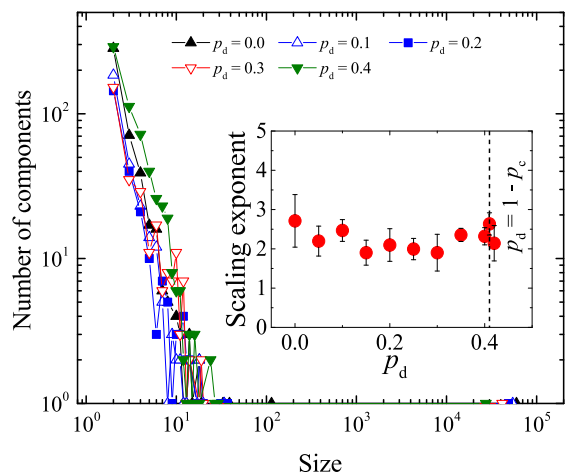}
	\caption{\label{fig7} (color online) An example of connected-component size distribution calculated for different fractions of defects. (inset) Computed scaling exponent of size distribution of connected-components over multiple realizations.}
\end{figure}

To evaluate small-world properties of the functional networks, we calculated the average shortest path length and the clustering coefficient of them for different values of lesion size. As is plotted versus the fraction of defects in Fig.~\ref{fig:NetParams(a)}, the average shortest path length ($l$) is reduced by increasing in $\pd$, but at the same time it is larger than the values obtained from an equivalent random network over the whole range below percolation limit.

Also, network clustering coefficient ($C$) is plotted in terms of $\pd$ in Fig.~\ref{fig:NetParams(b)}, showing that $C$ is larger than corresponding values of an equivalent random network in an order of magnitude up to the percolation limit. Based on the results of the Fig.~\ref{fig:NetParams}, it can be concluded that for the functional network, the characteristic length is finite and clustering coefficient is $\sim1$. The coexistence of these two characteristics means the functional network is small-world in the whole range under study. This description can be better quantified by defining the parameter of the small-worldness, commonly defined as \cite{Humphries2008}:
\begin{eqnarray}\label{eq.3}
S = \dfrac{C/C_\mathrm{ran}}{l/l_\mathrm{ran}},
\end{eqnarray}
which indicates that a higher clustering coefficient $C$ and lower average path length $l$ relative to the values of a corresponding random network ($C_\mathrm{ran}$ and $l_\mathrm{ran}$, respectively), leads to a network with higher small-world property. Fig.~\ref{fig:NetParams(c)} shows the resultant behavior of small-worldness of the functional network. Although this measure decreases monotonically with the amount of defects, it is at least $20$ times larger than that of an equivalent random network up to the vicinity of the percolation limit.

Another observation is that in the critical state of the Ising model, highly correlated sets of spins form densely connected components which their sizes also obey power-law behavior. The scaling of these distributions regarding different amount of defects is maintained and the corresponding exponents roughly remain in the same range (Fig.~\ref{fig7}). At the tail of the distribution function, the \red{traces of the largest components} are seen, which are more likely to contribute to the statistical results that were presented in our previous figures.

\begin{figure}
	\vspace{4mm}  
	\includegraphics[width=0.95\columnwidth]{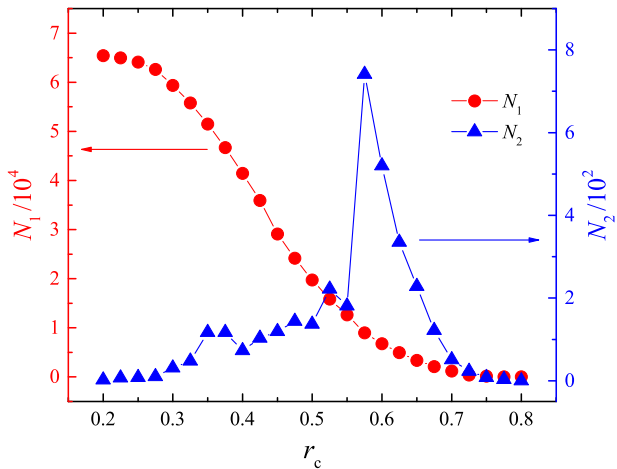}
	\caption{\label{fig8} \red{(color online) The number of nodes in the first two largest clusters of the functional network versus the values of $\rc$.}}
\end{figure}

It is clear that for small values of $\rc$ close to zero, the functional network will approach a trivial fully-connected network, and for large values of $\rc$, the network will be too sparse and the statistical parameters will not be reliable. So there is an optimal range from which the value of $\rc$ can be chosen to extract statistics (see also \cite{Zarepour2015} for more details). This range is generally determined by the sufficient number of nodes in the functional network and also the bounds on the distribution of correlations which varies according to $\pd$ [see Fig.~\ref{fig:N,E(b)}]. But more importantly, it is interesting to take into account the change of behavior of these networks as a function of $\rc$ in this range. Specifically, by definition, there is a percolation transition point across this range which can be identified by looking at the size of the second largest cluster relative to $\rc$ \cite{krioukov2012, gallos2012conundrum, gallos2012small}. As illustrated in Fig.~\ref{fig8}, we have observed a transition point at around $\rc\approx0.58$ for the clean system.\\
We have also recalculated all the relevant topological parameters previously discussed (in Figs.~\ref{fig:gamma}~and~\ref{fig:NetParams}) at this transition point, and observed that the network has the same qualitative behavior relative to $\pd$.\\

\section{Summary}
In the current work the behavior of the functional networks of 2D Ising model at the critical point against the defect on the lattice is investigated. Defects are introduced in the lattice by letting the spin of a fraction of nodes to zero. Our results showed that although the number of nodes in the functional networks decreases as the fraction of defect on the lattice grows, maintaining the criticality condition increases the number of edges (the edges across which the correlation exceeds the threshold $\rc$). We calculated the scaling exponent $\gamma$ of the degree distribution for the different defect ratios in the model. $\gamma$ decreases when lesion size $\pd$ increases and the system loses its hubs before the percolation limit roughly around $\pd\sim0.25$. We also studied the small-world properties of the functional network versus the degree of defect. The results showed that despite to smooth decrease of the small world measures of the network, the functional network remains small-world characterized by high clustering coefficient and low average shortest path length.

The Ising model has been extensively studied in the presence of impurities and defects, and changes in the critical behavior of the model have been inspected by computing the critical exponents \cite{mccoy1968theory1, *mccoy1969theory2, *mccoy1969theory3, *mccoy1970theory4}, but changes in the functional network of the commonly used models in the statistical mechanics with defects have not been studied before. Large-scale critical behavior of the brain activity has been explored by extracting the functional network from BOLD signal and the changes in characteristic parameters of the functional network are indicative of some neurodegenerative diseases \cite{Chialvo2010}. Moreover, evidence suggests that in neurodegenerative diseases like Alzheimer's, neurons that lose their function in the network gradually die over time before a symptom appears \cite{bredesen2006cell}. Therefore, there is a chance that the symptoms of the disease in the functional network be evaluated before other clinical measures. Study of the functional network of well-known models may reasonably increase our understanding of expected changes in the functional network due to defects in the structural network. Fortunately, multiple experimental evidences support our results about the functional network small-worldness and its relative parameters \cite{stam2006small, *zhao2012disrupted, *sanz2010loss}.

\begin{acknowledgments}
This research was supported by the Institute for Advanced Studies in Basic Sciences (Grant no. G2016IASBS12644), and by the Slovenian Research Agency (Grant nos. J1-7009 and P5-0027).
\end{acknowledgments}

\providecommand{\noopsort}[1]{}\providecommand{\singleletter}[1]{#1}%
\end{document}